\documentclass[useAMS,usenatbib]{mnras}

\newcommand{\mgii}{Mg~{\sc ii}}

\usepackage{graphicx}\usepackage{epsfig}\usepackage{epsf}
\usepackage{amsmath}\usepackage{amssymb}\usepackage{stfloats}
\title[UV ``God Rays'' from $\eta$ Car] {Ultraviolet Mg~II emission
  from fast neutral ejecta around Eta Carinae}

\author[Smith \& Morse]{Nathan Smith$^{1}$\thanks{E-mail:
    nathans@as.arizona.edu} and Jon A.\ Morse$^{2}$ \\
  $^{1}$Steward Observatory, University of Arizona, 933 N. Cherry
  Ave., Tucson, AZ 85721, USA \\
  $^{2}$BoldlyGo Institute, 1370 Broadway 5th Floor Suite 572, New York, NY 10018, USA}

\begin{document}

\pagerange{\pageref{firstpage}--\pageref{lastpage}} \pubyear{2019}
\maketitle
\label{firstpage}

\begin{abstract}
  We present the first images of the nebula around $\eta$~Carinae
  obtained with the Wide Field Camera 3 (WFC3) onboard the {\it Hubble
    Space Telescope} ({\it HST}), including an ultraviolet (UV) image in
  the F280N filter that traces \mgii\, emission, plus contemporaneous
  imaging in the F336W, F658N, and F126N filters that trace near-UV
  continuum, [N~{\sc ii}], and [Fe~{\sc ii}], respectively.
  The F336W and F658N images are consistent with previous images in
  these filters, and F126N shows that for the most part, [Fe~{\sc ii}]
  $\lambda$12567 traces clumpy shocked gas seen in [N~{\sc ii}].  The 
  F280N image, however, reveals \mgii\, emission from structures that 
  have not been seen in any previous line or continuum images of $\eta$~Carinae.  
  This image shows diffuse \mgii\, emission immediately outside the bipolar
  Homunculus nebula in all directions, but with the strongest emission
  concentrated over the poles.  The diffuse structure with prominent
  radial streaks, plus an anticorrelation with ionized tracers of
  clumpy shocked gas, leads us to suggest that this is primarily \mgii\,
  resonant scattering from unshocked, neutral atomic gas.  We
  discuss the implied structure and geometry of the \mgii\, emission,
  and its relation to the Homunculus lobes and various other complex
  nebular structures.  An order of magnitude estimate of the neutral gas mass 
  traced by \mgii\, is 0.02~$M_{\odot}$, with a corresponding kinetic energy 
  around 10$^{47}$ erg.  This may provide important constraints on polar mass 
  loss in the early phases of the Great Eruption.  We argue that
  the \mgii\, line may be an excellent tracer of significant reservoirs of
  freely expanding, unshocked, and otherwise invisible neutral atomic
  gas in a variety of stellar outflows.
\end{abstract}


\section{INTRODUCTION}

The massive evolved star $\eta$ Carinae is surrounded by a beautifully
complex system of nebulosity that seems to get more complicated and
interesting the closer we look.  The main components in images are the
prominent bipolar nebula known as the ``Homunculus''
\citep{gaviola50}, which is mostly a dusty reflection nebula, as well
as the more extended ragged splash of ionized condensations known as
the ``Outer Ejecta'' \citep{walborn76}.  The Outer Ejecta are
accompanied by a large shell seen in soft X-rays \citep{seward01},
indicating that fast ejecta are overtaking and shocking older, slower
material from previous mass loss.  The Outer Ejecta are N-rich
\citep{walborn76,davidson82}, while the degree of N enrichment is
greater inside the X-ray shell than outside of it \citep{sm04}.  These
nebular features represent the combined mass loss of a few energetic,
eruptive mass loss events over the past several hundred years,
including the ``Great Eruption'' in the mid-19th century and earlier
undocumented events \citep{morse01,kiminki16,smith17}.  Although
poorly understood, episodic mass loss may play an important role in
stellar evolution \citep{so06,smith14}, at least for a
subset of stars in interacting binaries \citep{st15,mojgan17}.

The eruption of $\eta$ Car and its resulting nebula are uniquely
valuable for studying episodic mass loss from massive stars because we
have the historical visual magnitudes of the 19th century outburst
\citep{herschel1847,sf11}, along with light-echo spectroscopy of
reflected light from the event
\citep{rest12,prieto14,smith18a,smith18b}, both of which can be
compared to contemporary extragalactic analogs of LBV eruptions
\citep{vdm12,smith+11}. Unlike extragalactic LBV eruptions, we can
also dissect the structural details of the remnant ejecta nebula that
is the mass-loss product of that previous event, and we can study the
current post-eruption state of the surviving star.

While the mass loss in the 19th century eruption has traditionally
been discussed as the result of a clumpy super-Eddington
radiation-driven wind \citep{shaviv00,owocki04,vanmarle08,vanmarle09},
some clues have hinted that an explosive mechanism may have been
important.  These clues include a high ratio of kinetic to radiated
energy \citep{smith03b}, the very thin walls and double-shell
structure of the Homunculus lobes \citep{smith06,smith13}, and some
very high observed velocities in the outer ejecta \citep{smith08}.
\citet{smith13} outlined how the nebula and historical light curve
could be reconciled with a model where the light curve was powered by
the shock interaction between fast explosive ejecta and slower
circumstellar material (CSM interaction), analogous to a scaled-down
version of a Type IIn supernova (SN).  More recently, spectroscopy of
light echoes from $\eta$ Car's eruption provide strong confirming
evidence of a shock-powered event \citep{smith18a,smith18b}, revealing
extremely high speed ejecta expanding at around 10,000 km s$^{-1}$ or
more, seen simultaneously with slower ejecta around 150 and 600 km
s$^{-1}$, and a time sequence of echo spectra that closely mimics the
spectral evolution seen in Type IIn SNe powered by CSM interaction.

These recent data imply that the current nebulosity
around $\eta$ Car may closely resemble a low-energy supernova
remnant (SNR).  Obviously the nature of the explosive event was
different, since core-collapse SNe are terminal explosions, whereas
$\eta$~Car's 19th century eruption was an incomplete, non-terminal
explosion that left behind a surviving very massive star.
Nevertheless, the current X-ray shell is quite similar to an SNR in
some ways \citep{seward01}, and the highly clumped, N-rich outer
ejecta can be understood as post-shock cooling features, analogous to
shocked CSM material like the N-rich quasi-stationary flocculi in Cas A
\citep{chev78,chev89}.  \citet{walborn76} pointed out the similarity
between $\eta$ Car's Outer Ejecta and these features in Cas A many
years ago.  Interestingly, the physical parameters of $\eta$~Car's
eruption also provide a surprisingly good match to predictions of
multiple massive shell collisions that occur in exotic events such as
pulsational pair instability SNe \citep{woosley17}.
Compared to traditional core-collapse SNe, non-terminal explosions
are, however, relatively unexplored and far less well understood;
detailed analysis of the structure, kinematics, and excitation of
nebulosity resulting from such an event may be quite valuable
for constraining future models.


The {\it Hubble Space Telescope} ({\it HST}) has had a tremendous
impact on our understanding of $\eta$~Carinae and its nebula.  The
first {\it HST} images of $\eta$~Carinae with WF/PC \citep{hester91},
WFPC2 \citep{morse98}, and ACS/HRC \citep{smith04a} each revealed new,
complex, and sometimes mysterious structural details in the
nebula.\footnote{Note that in this discussion we are referring to
  previous work on the extended nebula around $\eta$ Car.  In the
  interest of brevity, we do not refer to the vast literature
  concerning the central star, its immediate vicinity, its
  binary-induced multi-wavelength variability, or its colliding stellar
  winds.}  Multi-epoch images with {\it HST} also reveal detailed
photometric variability on small spatial scales in the nebula
\citep{smith00,smith04b,smith17}, as reflected starlight and line
emission respond to variations of the star during its 5.5 yr orbital
cycle \citep{damineli96} and long-term changes.  The unprecedented
angular resolution of {\it HST} has also enabled high-precision proper-motion 
measurements for the expanding Homunculus and other ejecta close
to the star \citep{currie96,morse01,dorland04,smith04a,wk12,smith17}. The
most recent results, with a baseline over nearly the entire post-COSTAR
lifetime of {\it HST}, yield a precise dynamical date of origin for the
Homunculus of 1847.1 $\pm$0.8 yr \citep{smith17}, while some features
close to the star and inside the Homunculus are younger
\citep{dorland04,smith04a}.  Similarly, {\it HST} images have been
used to measure the proper motions of the Outer Ejecta
\citep{morse01,kiminki16}, revealing a mix of dynamical ages that
indicate previous mass-loss episodes centuries before the Great
Eruption.  Overall, these {\it HST} images largely trace
starlight scattered by dust in the Homunculus (with some contribution
from line emission in various features), and line emission from
shock-ionized gas in the Outer Ejecta.

Additionally, both ground-based and {\it HST} spectra have provided
radial velocities that, when combined with structures in images,
helped to decipher the 3-D structure and kinematics of the Homunculus
and Outer Ejecta
\citep{ha92,weis99,weis01,weis12,davidson01,currie02,bish03,smith02,
  smith04,smith05,smith06,smith08,smith03a,teodoro08,wk12,steffen14,mehner16}.
High angular resolution mid-infrared (mid-IR) and submm images have
traced thermal emission from warm dust in the thin Homunculus lobes
and equatorial region \citep{chesneau05,smith03b,smith18c}, while
near-IR line emission from H$_2$
\citep{sd01,smith02,smith04,smith06,sf07,steffen14} and UV absorption
in the line of sight through the SE polar lobe
\citep{gull05,gull06,nielsen05,verner05} have probed the thin
molecular shell and dense atomic gas in the Homunculus walls.

A rarely noted feature that is directly relevant to the discussion
here is a polar bubble or shell immediately outside the Homunculus
south-east (SE) lobe seen in Ca~{\sc ii} HK absorption and He~{\sc i}
$\lambda$10830 absorption \citep{smith02,davidson01}, and in emission lines like
H$\alpha$ and [N~{\sc ii}] $\lambda$6584
\citep{currie02,smith03a,mehner16}, and near-IR [Fe~{\sc ii}]
\citep{smith02}.  There may be a corresponding bubble over the NW
polar lobe \citep{mehner16}, but it cannot be seen in
absorption. \citet{currie02} referred to the SE polar bubble as the
``ghost shell'', although it may be a more complex structure as
discussed below.  It appears related to -- but is not the same as -- features
seen in our new \mgii\, image.

Here we present the first UV images of $\eta$~Car in the light of emission from
the \mgii\, $\lambda\lambda 2796, 2803$ resonant doublet
taken through the F280N filter with the {\it HST} Wide  
Field Camera 3 (WFC3).  UV imaging at similar wavelengths in 
broad continuum filters (F220W and F250W) obtained with the ACS High 
Resolution Channel (HRC) was reported previously \citep{smith04a,smith04b}, 
but had shorter exposures only tracing the bright Homunculus, dominated 
by dust-scattered starlight plus some diffuse line emission that creates 
a ``Purple Haze'' close to the star.  The new \mgii\, images presented 
here are deeper, and the UV line emission reveals structures not traced 
by any previous diagnostic.  Section 2 presents the new {\it HST}/WFC3 
observations, Section 3 briefly lists the main results of the imaging, 
and Section 4 includes an interpretation of various features, an 
analysis of the \mgii\, emission, and discusses the structure of the 
Homunculus and Outer Ejecta in context with several other multi-wavelength 
diagnostics from the literature.

\begin{table}\begin{center}\begin{minipage}{3.2in}
\caption{WFC3 UVIS and IR Channel Observation Log$^{a}$}
\scriptsize
\begin{tabular}{@{}lcccl}\hline\hline
UT Date &Filter  &$\lambda_{peak}$/Width &Diagnostic &Total \\
(2018) & \, & (\AA) & (Ejecta) & Exposure \\
\hline
05 Mar  &F280N   &2798/43     &Mg {\sc ii}  &7860s  \\
06 Mar  &F126N   &12590/152    &[Fe {\sc ii}] &1997s  \\
25 Jul  &F280N   &2798/43    &Mg {\sc ii} &6, 40 s \\
25 Jul  &F336W   &3375/511    &NUV, [N~{\sc i}] &1, 16, 240s  \\
25 Jul  &F658N   &6585/28    &[N~{\sc ii}] &1, 16, 540s  \\
\hline
\end{tabular}\end{minipage}\end{center}
$^{a}$ \scriptsize Programs GO-15289 and GO-15596.
\end{table}

\begin{figure*}
\includegraphics[width=6.5in]{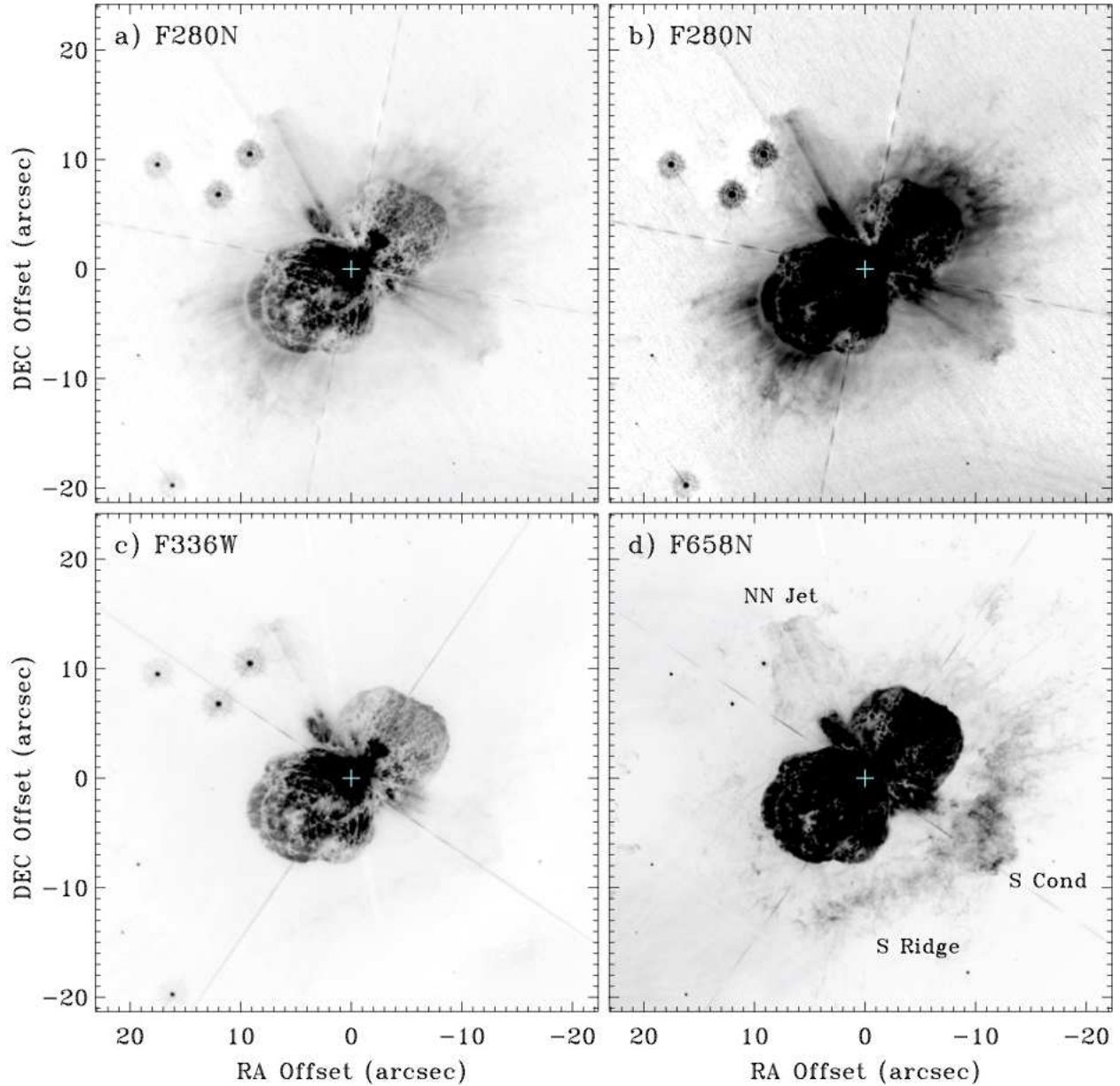}
\caption{Comparison of new deconvolved 
  near-UV {\it HST}/WFC3 images of $\eta$ Car in the F280N \mgii\, filter (a,b),
  the F336W near-UV continuum filter (c), and the F658N [N~{\sc ii}] 
  $\lambda$6584 filter (d).  Panels $a$ and $b$ show the same F280N 
  image with two different intensity stretches, with Panel $a$ scaled 
  to show brighter features for comparison with the F336W continuum image, 
  and with Panel $b$ scaled to show the faintest detected Mg~{\sc ii} emission.}
\label{fig:img}
\end{figure*}

\begin{figure*}
\includegraphics[width=5.5in]{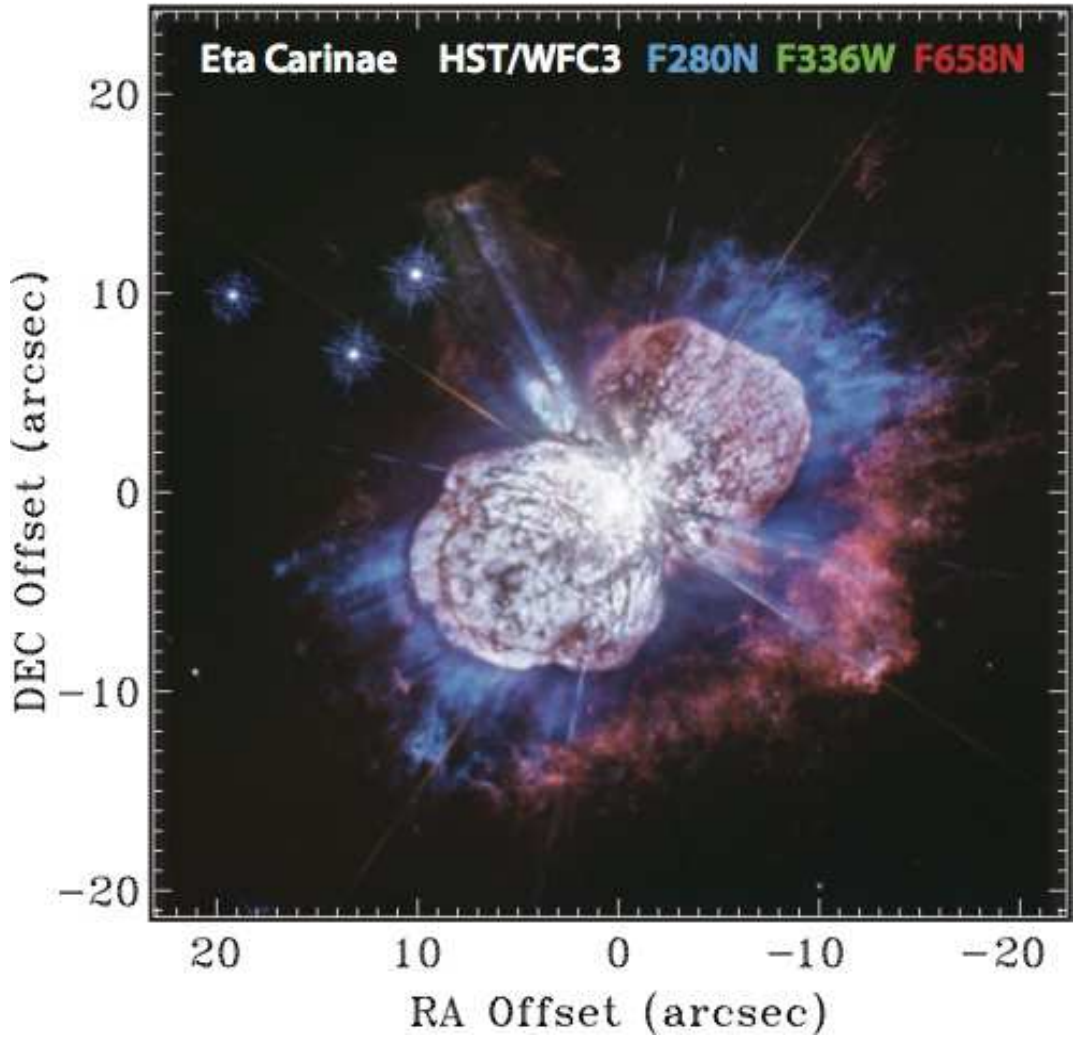}
\caption{Color {\it HST}/WFC3 image of $\eta$ Car with F280N in blue,
  F336W in green, and F658N in red.}
\label{fig:color}
\end{figure*}

\begin{figure*}
\includegraphics[width=6.5in]{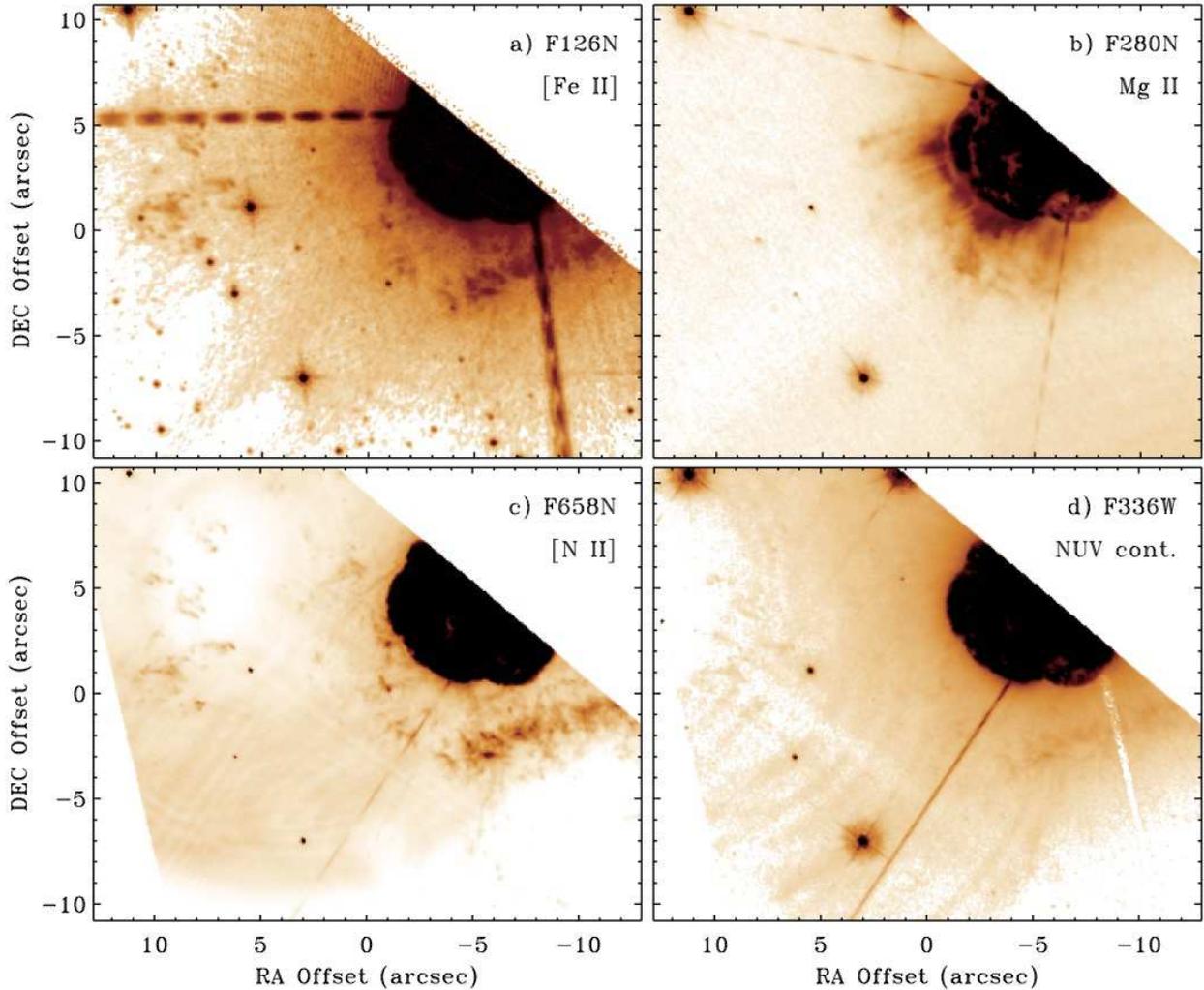}
\caption{Images aligned to the [Fe~{\sc ii}] F126N image, which
  covered only part of the region around $\eta$~Car in order to avoid
  the extremely bright central star.  (a) The
  F126N image, sampling [Fe~{\sc ii}] $\lambda$12567, (b) the F280N
  image, (c) the F658N image including [N~{\sc ii}] $\lambda$6584 and
  some redshifted H$\alpha$, and (d) the F336W image, which includes
  mostly near-UV continuum, but includes some line emission from
  [N~{\sc i}] $\lambda$3466 and various fainter Fe~{\sc ii} lines.  The bright
  background in F658N and F336W is due to ghost images of the central
  star.  Positional offsets in arcsec are from an arbitrary point at
  the center of the image.}
\label{fig:fe2}
\end{figure*}

\begin{figure*}
\includegraphics[width=6.1in]{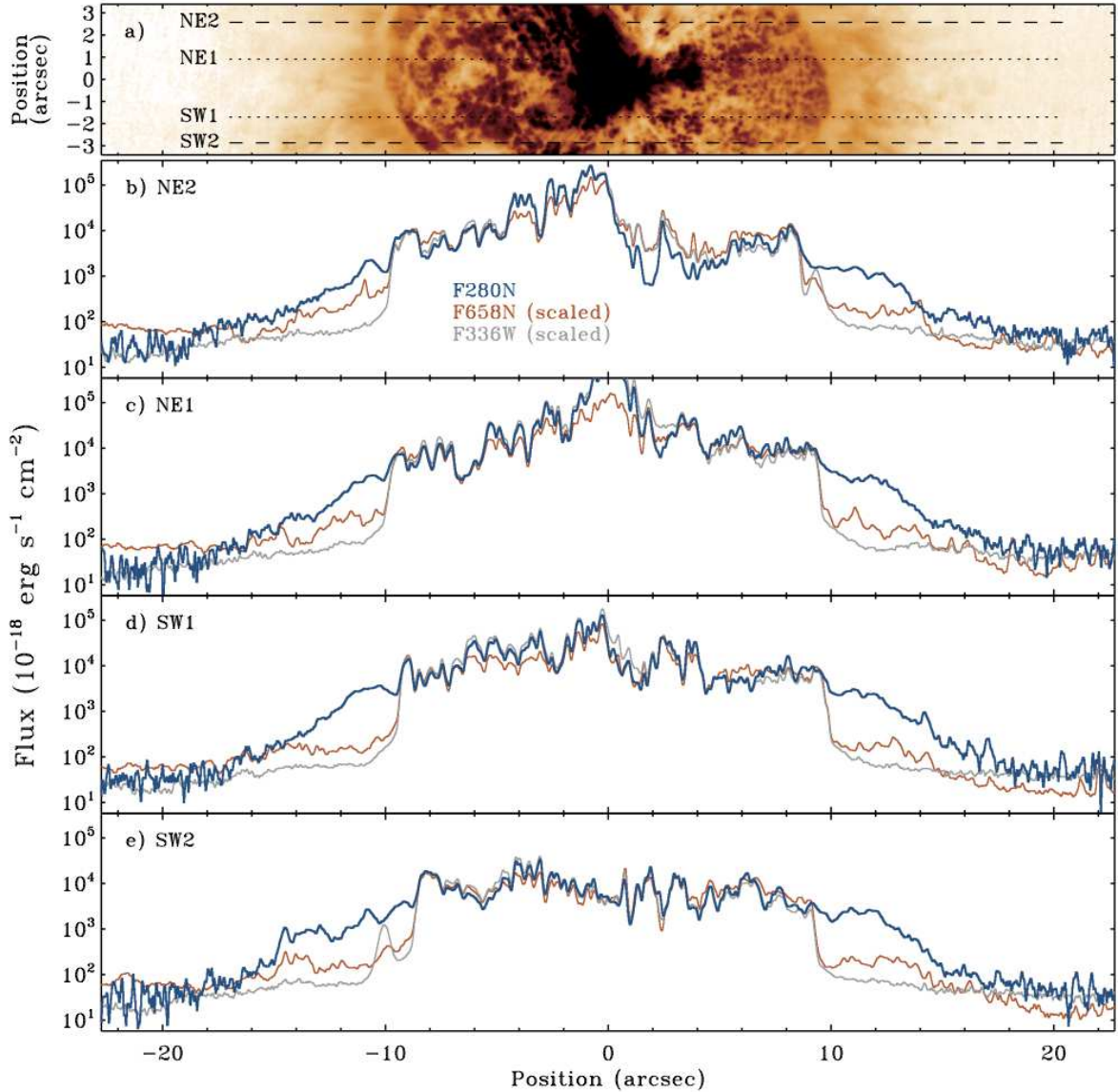}
\caption{Intensity scans across the Homunculus in WFC3/UVIS images.
  Top panel ($a$) shows a portion of the F280N image of the
  Homunculus, rotated clockwise by 40$^{\circ}$ so that the polar axis
  is horizontal.  Positions of four scans across the nebula are
  indicated (northeast 1 and 2, southwest 1 and 2).  Panels $b$
  (northeast 2), $c$ (northeast 1), $d$ (southwest 1), and $e$
  (southwest 2) show intensity scans along the positions indicated in
  panel $a$.  Each of these scans averaged the flux across 3 spatial
  pixels, or 0$\farcs$07.  Blue is the line flux in the F280N filter.
  Orange and grey show scaled intensities of F658N and F336W,
  respectively, where their relative intensity has been adjusted
  roughly to match the F280N flux of the polar lobes of the Homunculus
  for comparison.}
\label{fig:scan}
\end{figure*}

\begin{figure*}
\includegraphics[width=6.4in]{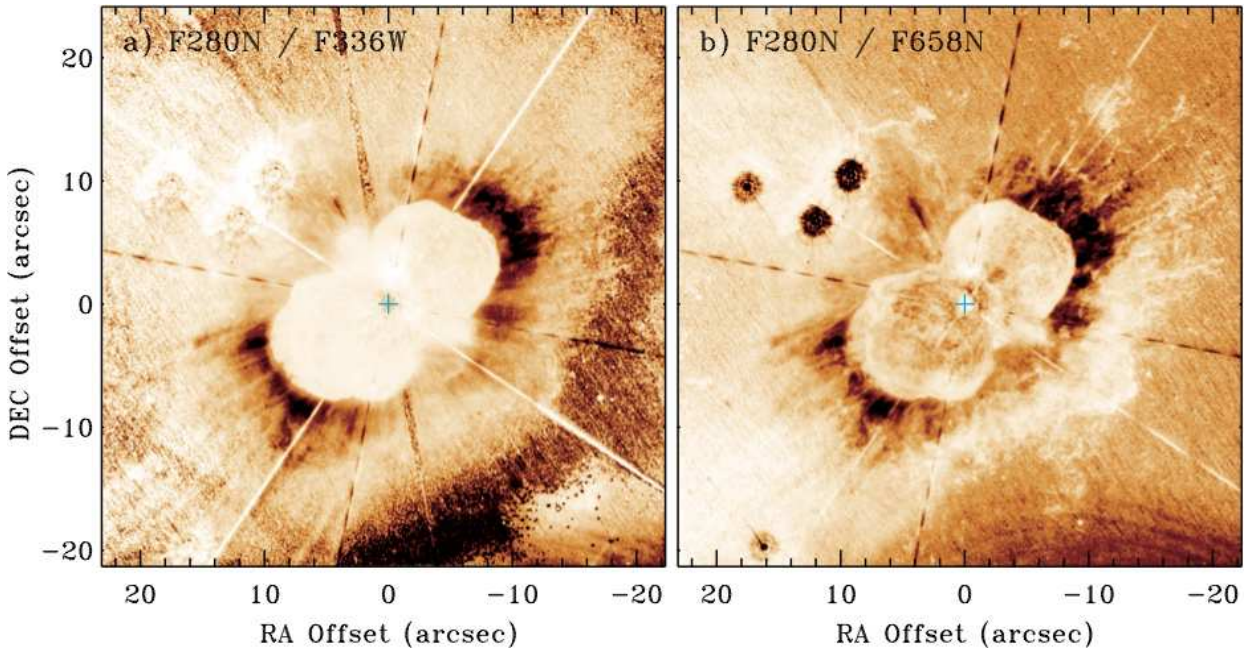}
\caption{Ratio images. (a) Relative flux ratio of F280N / F336W.
  Features that are relatively bright in F280N are dark, and the near
  UV continuum or [N~{\sc i}] emission is light.  (b) Ratio image of
  F280N / F658N, where regions with relatively strong Mg~{\sc ii}
  emission are dark, and locations of strong [N~{\sc ii}] emission are
  light. The lighter regions of diffuse background at the left
  portions of each panel are due mostly to excess flux from ghost
  images of the bright central star in the F336W and F658N images.}
\label{fig:ratio}
\end{figure*}

\begin{figure*}
\includegraphics[width=6.5in]{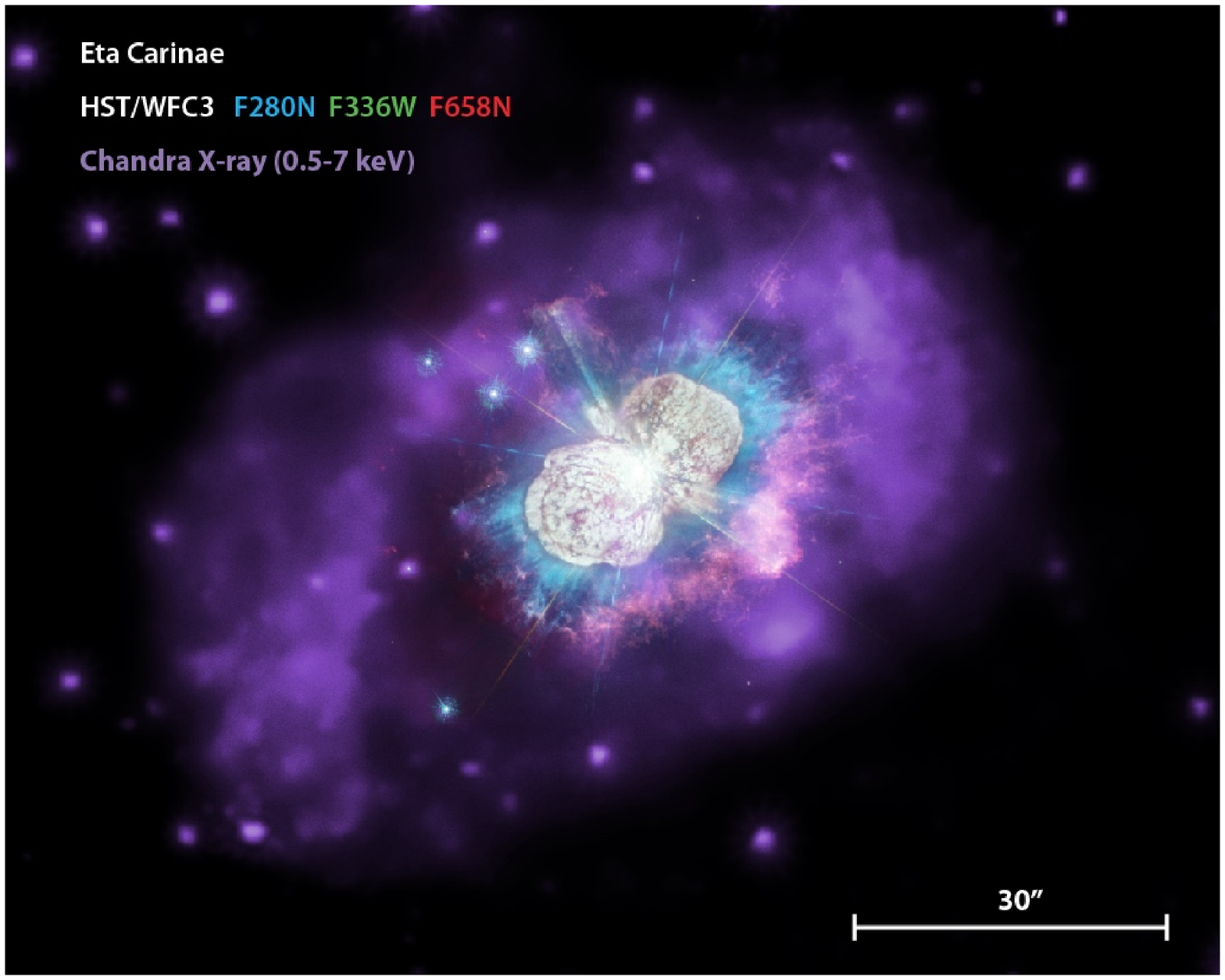}
\caption{The adaptively smoothed 0.5-7 keV X-ray image of the Outer Ejecta 
as seen by {\it Chandra} ACIS-S is shown in purple.  Processing and analysis 
of similar {\it Chandra} imaging of $\eta$ Car's Outer Ejecta have been 
discussed in detail previously \citep{seward01,corcoran04,hamaguchi14}.  
The 3-color WFC3/UVIS image from Figure~\ref{fig:color} is superposed.}
\label{fig:xray}
\end{figure*}

\section{OBSERVATIONS}

$\eta$ Car was observed through the UVIS and IR channels of WFC3 in two 
observing epochs.  The initial observations ({\it HST} program ID 15289) 
occurred in March 2018 through the UVIS F280N (Mg~{\sc ii}) filter and 
IR F126N ([Fe~{\sc ii}]) and F130N (IR continuum) filters. 

We used the standard STSDAS pipeline image reduction process for all images, which
performed bias-subtraction, flat-fielding, removal of
cosmic rays, geometric distortion correction and flux calibration.
Small offsets/dithers were used during the observations to 
recognize and correct for bad pixels on the UVIS CCD detectors 
and to limit the number and bleeding of saturated CCD pixels caused by the bright central star. 
Ultimately, twelve 655 sec F280N exposures, for a total exposure time of 7860 secs, 
were combined by the DRIZZLE task into a single Mg~{\sc ii} image. 
This deep image was captured to trace emission from the fast-moving ejecta 
outside of the Homunculus; 
however, as a result the inner $\sim$1 arcsecond around the central star was 
saturated in the final processed image.

The IR channel observations were programmed to avoid placing the central star on the detector. 
Using different roll angles and spatial offsets, the central star was moved 
sequentially off the edges of the square detector so that the rims of the 
Homunculus lobes and outer ejecta could be observed. 
Unfortunately, two of the four orientations appear to have experienced guiding problems, 
so that F126N and F130N mosaics of the entire region around the Homunculus were not possible. 
In this study, we focus on the F126N ([Fe~{\sc ii}]) image in the orientation that 
captured the rim of the SE lobe, S Ridge and E Condensations, using the terminology 
of \citet{walborn78}, and compare this to the emission observed through other filters. 

The purpose of comparing the UV \mgii\ and near-IR [Fe~{\sc ii}] is that these ions 
trace similar temperatures in post-shock gaseous structures. 
These lines appear in radiative shock models \citep{hartigan87,hartigan04}, along with
other common visible emission lines (H$\alpha$, [N~{\sc ii}], [S~{\sc ii}], etc.)
from different temperature regimes,
and may be compared to indicate local density, extinction and/or abundance variations.
Also, the near-UV imaging offers the highest spatial resolution full-field WFC3 emission-line imaging.

As discussed in more detail in Sec.~3, there is significant overlap between 
the \mgii\, and [N~{\sc ii}] emission, particularly in bright [N~{\sc ii}] outer ejecta knots
such as the S Condensation. However, those are not the brightest features in the \mgii\, image,
which instead reveals a host of hitherto unseen filaments and structures related to
both bipolar and equatorial structures. The \mgii\, emission largely complements -- that is, lies
adjacent to, often along contiguous structures -- the [N~{\sc ii}] emission. This intriguing 
relationship was evident even when comparing the new F280N image 
with {\it HST} WFPC2 F658N images from the 1990s, despite arcsecond-scale 
proper motions of the bipolar lobes and outer debris. In order to make detailed comparisons at
the scale of small knots and filaments, we therefore requested and were approved for 1 orbit of 
Director's Discretionary Time (program ID 15596) to obtain contemporaneous WFC3 [N~{\sc ii}] 
(F658N) and blue continuum (F336W) observations, which were filters used in past 
imaging {\it HST} imaging campaigns. We also acquired short duration \mgii\, (F280N) images to
mitigate the saturated pixels near the central star in the original program. 

Sequences of images with different exposure times were obtained 
in July 2018 through the F280N, F336W and F658N filters in order 
to cover the large dynamic range between the $\eta$ Car
central star and faint filaments in the outer debris 
(see Table 1). In order to fit the exposure sequences into
a single WFC3 buffer dump to the {\it HST} solid state recorder, 
we used 1024$\times$1024 subarrays for the short and medium 
exposures -- since we were only interested in the pixels around 
the central star -- and then 2048$\times$2048 subarrays for the longest 
F336W and F658N exposures. Saturated pixels near the central 
star and associated bleeding in longer exposures were
masked out, and pixel flux values from shorter exposure images 
were scaled and substituted. Even though the 2048$\times$2048 subarray
used for the long exposures was only a quarter of the 
full WFC3 field of view, the images covered most of the 
important Outer Ejecta and, at 40 milli-arcseconds per pixel, 
had essentially the same field of view as previous WFPC2 WF 
camera images (approximately 800x800 at $\sim$100 
milli-arcseconds per pixel) but with better than PC resolution. 
Finally, we deconvolved each
image conservatively for 3 iterations using the Lucy-Richardson 
algorithm provided in STSDAS. The images were rotated and
aligned to our historical dataset of WFPC2 images
using IRAF{\footnote{IRAF is distributed by 
the National Optical Astronomy Observatories, which is operated 
by the Association of Universities for Research in Astronomy,
Inc.~(AURA) under cooperative agreement with the National 
Science Foundation. The Space Telescope Science Data Analysis 
System (STSDAS) is distributed by the Space Telescope Science 
Institute.}} tasks and about seven stars common to all
images as tie points.

Table 1 summarizes the observations used in this study, 
indicating the filters, exposure times, and principal 
contributors to the observed features. 
We note that every image through all the filters contains some 
sort of internal reflection ghost and/or scattered light from the
bright central star, 
which vary as a function of {\it HST} roll angle and where the 
central star was placed in the field of view. 
Such artifacts limit how faint we can see in certain regions, 
and we therefore focus on the 
Homunculus and brighter Outer Ejecta emission-line features.
We also note detectable fractional-pixel proper motions of the fast-moving debris between 
the March 2018 exposures through the F280N filter and those from July 
2018 through F336W and F658N. The motions are barely discernible 
when blinking the images in an image display, but we nonetheless
avoid over-interpreting very small-scale differences until we can 
obtain images within the same day or week in a future campaign.

It is also important to keep in mind, particularly for the narrowband filters,
the limited and varying velocity ranges that are recorded in the images.
The F658N filter has the narrowest width (see Table 1), 
covering a velocity range in [N~{\sc ii}]$\lambda6583$ emission of about 
$\pm$650 km~s$^{-1}$, plus redshifted H$\alpha$ emission from about +350
to +1600 km~s$^{-1}$. The F280N filter covers a larger velocity range 
in \mgii\ emission of about $\pm$2300 km~s$^{-1}$, while the near-IR F126N filter
covers $\pm$1800 km~s$^{-1}$ in [Fe~{\sc ii}]. Given these velocity range differences, 
it is possible that there is some missing emission when comparing the morphologies 
in the different diagnostic features. Nevertheless, given the similarities between
the [Fe~{\sc ii}] and [N~{\sc ii}] emission features, and how distinct the \mgii\ features 
between the Homunculus lobes and outer, [N~{\sc ii}]-bright ejecta generally are
(see discussion below), we believe that the observed emission represents real
distributions of the neutral and ionized gas, and are not artifacts of the 
different filter widths. Future proper-motion and spectroscopic observations could definitively verify this.

\section{RESULTS}

Various features in the new WFC3 images are discussed in the following subsections.  Figure~\ref{fig:img} shows the deconvolved WFC3-UVIS images in the F280N (two intensity ranges), F336W, and F658N filters.  Figure~\ref{fig:color} presents a 3-color composite of these same images with F280N in blue, F336W in green, and F658N in red.  Figure~\ref{fig:fe2} shows the portion of the Outer Ejecta imaged in [Fe~{\sc ii}] emission in the F126N filter of WFC3-IR, as well as aligned portions of the same three WFC3-UVIS images as in previous figures.  Figure~\ref{fig:scan} shows intensity scans crossing the Homunculus at several positions parallel to  the polar axis in the F280N (blue), F336W (gray), and F658N (orange) filters.  Figure~\ref{fig:ratio} shows images representing the intensity ratio of F280N/F336W and F280N/F658N, highlighting the differences between Mg~{\sc ii} emission and the other lines.  Finally, Figure~\ref{fig:xray} illustrates the spatial relationship between the 0.5-7 keV X-ray emission from the Outer Ejecta seen by {\it Chandra}, shown in purple, and the same 3-color WFC3-UVIS image from Figure~\ref{fig:color}.

\subsection{F280N Mg~II Emission}

In the F280N filter, the main part of the bipolar Homunculus nebula is dominated by near-UV dust-scattered starlight, just as it is in most broad band UV/optical filters, and very similar to the near-UV scattered continuum in the F336W filter (compare Figures~\ref{fig:img}a and \ref{fig:img}c).  However, F280N also reveals \mgii\, emission structures immediately outside the Homunculus that are either absent or far less pronounced at any other wavelength range that we are aware of (Figure~\ref{fig:img}; compare panels $a$ \& $b$ to $c$ \& $d$).  Previous UV spectra confirm that \mgii\, emission dominates the flux in the F280N filter at locations in the Outer Ejecta near the S~Condensation \citep{davidson82,dufour99}.

In general, the brightest F280N emission outside the Homunculus appears anticorrelated with [N~{\sc ii}] emission in the F658N filter.  This is best appreciated in the color image in Figure~\ref{fig:color} where F280N emission (blue) dominates at most locations immediately outside the main polar lobes of the Homunculus, whereas F658N emission (red) is stronger in a shell at larger separation from the star.  Figure~\ref{fig:color} gives the impression that \mgii\, emission fills the cavity inside the [N~{\sc ii}]-emitting shell.
While a subset of the features seen in F280N can also be seen in F336W and/or F658N, 
the morphology and intensity distribution is different between these filters, 
and many structures are only seen in one filter. 

Spatial intensity scans of F280N, F336W, and F658N flux are shown in Figure~\ref{fig:scan}, showing that the strength of \mgii\, emission immediately outside the Homunculus is almost as bright as the scattered near-UV continuum of the polar lobes, whereas the F336W and F658N flux drops precipitously at the outer boundary of the polar lobes.  F658N (in some directions) brightens at larger separation, whereas the F280N flux tends to decline smoothly with increasing separation from the star.  

Ratio images of F280N/F336W and F280N/F658N are shown in Figures~\ref{fig:ratio}a and \ref{fig:ratio}b, respectively.  These highlight structures that are particularly enhanced in \mgii\, line emission but not in other tracers, because they divide out the near-UV or red continuum scattered light, as well as the H$\alpha$ and [N~{\sc ii}] line emission.  While \mgii\, emission is seen from a blue halo encircling the Homunculus in Figure~\ref{fig:color}, it is also evident from Figure~\ref{fig:ratio} that the \mgii\, emission is most pronounced over the poles.

Strong \mgii\, emission, combined with a lack of H$\alpha$ or [N~{\sc ii}], indicates that the F280N filter in regions immediately outside the Homunculus (blue in Fig.\ref{fig:color}, black in Fig.\ref{fig:ratio}) is tracing predominantly neutral atomic gas (i.e., neutral H, but low-ionization metals), because Mg$^0$ is ionized to Mg$^+$ at 7.6 eV, whereas Mg$^+$ is ionized at 15 eV.  Below in Section 4.1 we discuss why we favor an interpretation of resonant scattering for the \mgii\, line production rather than post-shock cooling emission.
Furthermore, as previously noted, given that the \mgii\ emission generally
lies spatially between the Homunculus lobes and outer ejecta bright in [N~{\sc ii}] --
such as the S Condensation -- we do not believe that the larger velocity range 
transmitted through the F280N filter is responsible for the morphological differences,
and indeed indicates that the \mgii\ emission (neutral gas) may be confined to lower
velocities than the ionized features seen in [N~{\sc ii}].

\subsection{F336W and F658N}

New images in the F336W and F658N filters (Figs.~\ref{fig:img}c and 
\ref{fig:img}d) show qualitatively the same structure as seen in previous 
epochs of imaging, as described already by \citet{morse98}
and other authors.  We do not analyze these images further, except to
compare them with new images in the F280N and F126N filters.  For a
detailed comparison with F280N emission, new images in these filters
were needed because of the rapid expansion of the nebula.

The flux detected in the broad F336W filter is dominated by near-UV
starlight scattered by dust grains in the Homunculus, but there is also a minor contribution from
emission lines of low-ionization metals.  In the Outer Ejecta, there is no strong contribution from
high ionization lines like Ne~{\sc iii} or Ne~{\sc v} as in some supernova remnants \citep{sd95,blair00}.
Referring to the archival {\it HST} FOS UV spectrum of the 
S Condensation (see \citealt{dufour99}),
the strongest emission line in the F336W bandpass is [N~{\sc i}] $\lambda3466$.
Such emission is not surprising given the N-rich abundances in the Outer Ejecta, and the wide range of ionization \citep{davidson82,sm04}.
This [N~{\sc i}] emission is faint, and it appears to track the [Fe~{\sc ii}] and [N~{\sc ii}] emission (see below).

The narrow F658N filter is also dominated by scattered continuum
starlight in the bipolar lobes of the Homunculus (or perhaps more accurately, a combination of scattered continuum plus dust-scattered and redshifted H$\alpha$ from the central star's wind; see \citealt{smith03a}), but is primarily tracing intrinsic [N~{\sc ii}] $\lambda$6584 emission and 
a small amount of redshifted H$\alpha$ recombination emission in
the Outer Ejecta.  In the Outer Ejecta, this emission traces gas that
has been collisionally excited and ionized by a shock, which is rapidly 
cooling and fragmenting.  This is why the F658N image of the Outer Ejecta has a morphology consisting of a complex arrangement of strings, clumps, shells, filaments, etc. \citep{morse98,sm04,weis99,weis01}.

\subsection{F126N [Fe~II] Emission}

Near-IR emission from [Fe~{\sc ii}] (especially the pair of strong
lines at 12567 and 16435 \AA) is seen from the shell nebulae around
several LBVs, including $\eta$ Carinae, P Cygni, and others
\citep{ha92,smith02,smith02b,sh06}.  These lines are thought to be
strong in LBVs because the shells are composed of dense,
low-ionization gas \citep{smith02b,sh06}.  LBVs are strong sources of
non-ionizing Balmer continuum UV radiation, but their dense winds can
absorb much of the Lyman continuum flux.

Figure~\ref{fig:fe2} shows a WFC3/IR-channel image in the near-IR F126N
filter, which samples [Fe~{\sc ii}] $\lambda$12567.  The field of view
is offset to the SE from the star to avoid the bright central star and
the core of the Homunculus; even the SE lobe is saturated in this
image.  Three other pointings (to the NE, NW, and SW of the star) were planned,
but the resulting exposures suffered from tracking problems and other issues 
that compromised the quality of the images (as noted above in Section 2).

Figure~\ref{fig:fe2} also shows images aligned to this F126N image in
three other filters: F280N (\mgii), F658N (H$\alpha$ and [N~{\sc ii}]
$\lambda$6583), and F336W (near-UV continuum plus some weaker emission
from [N~{\sc i}] $\lambda$3468 and several Fe~{\sc ii} lines).  From
this comparison, it seems clear that the [Fe~{\sc ii}] line emission is
tracing mostly the same gas that glows brightly in the [N~{\sc ii}]
line in F658N.  The E condensations and the dense knots that compose
the S Ridge are seen clearly in both F126N and F658N.  Since the
N-rich knots in these Outer Ejecta exhibit a wide range of ionization
in lines from [N~{\sc i}] to N~{\sc v} \citep{davidson82}, they are tracing
dense condensations that are rapidly cooling after being ionized by a
shock.

On the other hand, the emission in the F280N filter is tracing very
different emitting structures from the F658N and F126N filters.
Although some of the F280N emission outside the SE lobe overlaps with
features seen in F658N, these emitting structures have a very
different morphology and radial extent.  The F280N emission arises
from an entirely different set of gaseous structures, or perhaps a
combination of these different structures and some emission from the
same knots seen in F658N and F126N.  (Spectra of the \mgii\, line are needed to
determine if the radial velocities are distinct from the velocities of
[N~{\sc ii}] and [Fe~{\sc ii}] emission.)  Whereas the F658N and F126N 
filters trace a series of dense clumps that are separated from the Homunculus, 
the F280N emission is more diffuse, and often exhibits radial streaks 
immediately outside the polar lobes of the Homunculus.  The
F280N structures are therefore both morphologically and spatially
distinct from the shocked [N~{\sc ii}] and [Fe~{\sc ii}] clumps.  This
is one of the main reasons we favor the interpretation that the
emission in the F280N filter is resonant scattering of \mgii\, from
freely expanding, low-ionization, unshocked ejecta outside the
Homunculus (see below).

\section{DISCUSSION}

\subsection{Mg~II Resonant Scattering}

We propose that the bulk of the \mgii\, emission detected in the F280N filter that is seen outside the Homunculus but inside the N-rich shell of Outer Ejecta arises predominantly from resonant scattering in the \mgii\, $\lambda\lambda$2796,2803 doublet.  
Resonant scattering traces warm neutral atomic gas, rather than recombination emission from ionized and cooling post-shock gas.  
Although \mgii\, line intensity ratios, line profiles, and radial velocities in spectra will be helpful to confirm this, our conjecture of resonant scattering based on images alone relies upon 3 key observed properties:

1.  The structures immediately outside the Homunculus are unique to images in the F280N filter; these structures are not seen in any other continuum or emission line images of $\eta$ Car that we are aware of.  Since they are not seen in the adjacent broad F336W filter, they are not due to UV continuum scattered light and must be \mgii\, line emission.  However, gas that emits \mgii\, through recombination or collisional excitation because it is photoionized or shock excited would also be seen in other emission-line diagnostics like H$\alpha$, [N~{\sc ii}], or [Fe~{\sc ii}].  Resonant scattering is a mechanism to preferentially cause \mgii\, emission. 

2.  The morphology of the F280N features shows radial streaks and a smoother overall distribution than other line-emitting features in the Outer Ejecta.  Long, smooth, radial streaks may arise naturally if they are tracing beams of scattered starlight escaping through holes and cracks with low optical depth, in between dust clumps.  
Indeed, the polar lobes of the Homunculus present a complicated network of dust clumps and filaments \citep{morse98}.  Shadows cast by these clumps contribute to the contrast of the streaks, with dark lanes projecting out radially.  
Similarly, long wavelength imaging of the equatorial belt inside the Homunculus reveals a clumpy, disrupted torus with holes through which starlight can escape \citep{chesneau05,smith02c,smith18c,smith03b}.  
This may help account for the prominent radial streaks that connect the Homunculus equatorial skirt to shocked clumps in the NN Jet and S Condensation.  
However, since these radial streaks are absent, less prominent, or have different intensity distributions in continuum images, they cannot be due to dust scattering.  Line resonance scattering is needed.    
In stark contrast, emission features in the Outer Ejecta seen in other lines like [N~{\sc ii}] and [Fe~{\sc ii}] --- which {\it are} expected to trace post-shock cooling emission --- do not show a smooth and radially streaked morphology.  
Instead, these lines trace a highly complex network of small-scale clumps and filaments that might arise naturally from rapid radiative cooling in dense post-shock gas. 

3. The location of the most prominent F280N emission appears as a diffuse halo sandwiched in between a thin shell of cold, dense, molecular gas and dust in the Homunculus \citep{smith02,smith06,smith03b} at its inner boundary, and the shell of hot, shock-ionized outer ejecta \citep{walborn76,davidson82,sm04,weis99,weis01} at its outer boundary.  
This is an intermediate ionization transition zone where one might naturally expect to find warm and predominantly neutral atomic gas, well-suited to produce \mgii\, resonance scattering if it is bathed in non-ionizing radiation.  
Indeed, this region is exposed to strong near-UV Balmer continuum
(non-ionizing radiation) from $\eta$~Car at locations where starlight can escape through cracks in the clumpy Homunculus.  
This is non-ionizing Balmer continuum starlight that escapes, because essentially all the Lyman continuum is choked off at much smaller radii by $\eta$~Car's extremely dense
stellar wind \citep{hillier06,smith03a}.  This escaping near-UV starlight can ionize Mg$^0$ to Mg$^+$ (7.6 eV), but because there are no Lyman continuum photons, it cannot ionize the gas to Mg$^{++}$ (requiring 15 eV).
This makes it highly probable that near-UV starlight escaping through holes in the Homunculus will encounter Mg$^+$ atoms in the ground state, and emit the streaked \mgii\, $\lambda\lambda$2796,2803 that we see.

The \mgii\, $\lambda\lambda$2796,2803 resonant doublet ($3s ^2S-3p ^2P^0$) is magnesium's analog of the Ca~{\sc ii} HK lines.  Ca~{\sc ii} HK emission may be weaker or absent in this same region, however, because of the lower Ca abundance (relative to Mg) at solar metallicity, and because Ca$^+$ can be ionized to Ca$^{++}$ by near-UV photons above 11.8 eV but below 13.6 eV, which may escape $\eta$~Car's stellar wind and the Homunculus.  
It is interesting to note, however, that Ca~{\sc ii} HK has been detected in long-slit spectra, but seen in {\it absorption} arising from a thin shell of neutral gas immediately outside the SE polar lobe of the Homunculus \citep{davidson01,smith02}.  
This Ca~{\sc ii} structure in absorption is not the same as the \mgii\, emission structures discussed here, but it may be related (see Section 4.3 below).

Generally, \mgii\, emission is seen in all directions immediately outside the Homunculus and inside the N-rich shell that includes the S Ridge, S~Condensation, and NN~Jet, but the \mgii\, emission is brightest over the poles, while the spikes are most dramatic in the equatorial skirt.  
Regions with strong \mgii\, emission also tend to be locations where [N~{\sc ii}] emission is relatively weak.  Some of the detailed \mgii\, features are discussed below.

\subsection{Structure of Mg~II over the poles and the Ghost Shell}


Much of the F280N flux is concentrated in dramatic radial structures in the gas beyond the edges of the bipolar Homunculus lobes (Figure~\ref{fig:ratio}).  The \mgii\, emission just outside the SE (approaching) lobe bears the hallmarks of so-called ``God rays'' --- such as seen at the edges of terrestrial clouds or mountains --- due to shadows cast by opaque dust features on the lobe surface. This important phenomenon seems to isolate soft UV emission from the central star as the source of exciting resonantly scattered photons (see above).  The \mgii\, emission structures outside the NW (receding) lobe may also result from opaque dust features on that lobe, though the morphology is more filamentary, reminiscent of the spiked hair of a troll doll.

A second interesting feature is the dark rim right at the edge of the SE lobe where there is a $\sim$0.5{\arcsec}-wide deficit of \mgii\, emission.  There may be an isolated example of a similar dark edge at one position at the bottom of the NW lobe, but in general there is no similarly prevalent dark feature around the edge of the NW lobe. The origin of this dark edge of the SE lobe and the general lack of a similar feature at the edge of the NW lobe is unclear.  One possibility is that it is either a shadow or absorption caused by high optical depths at the limb-darkened edge of a thin and dense gas shell immediately outside the Homunculus lobes.  Alternatively, \mgii\, emission could be absent at this location, in principle, because Mg atoms in a thin layer on a skin outside the Homunculus lobes are either predominantly ionized to Mg$^{++}$ or have recombined to be mostly neutral Mg$^0$.

Although the specific features traced by \mgii\, emission outside the polar lobes have not been seen in any previous imaging of $\eta$ Car, some observations have suggested the existence of a possibly related polar shell or bubble outside the SE polar lobe.  
Optical and IR long-slit spectra have revealed blueshifted absorption from a thin (unresolved velocity width much less than the expansion speed of several hundred km s$^{-1}$) 
polar structure outside the Homunculus SE polar lobe seen in Ca~{\sc ii} HK absorption and He~{\sc i} $\lambda$10830 absorption \citep{davidson01,smith02}, 
as well as narrow H$\alpha$ and [N~{\sc ii}] emission \citep{currie02,smith03a,mehner16} 
and near-IR [Fe~{\sc ii}] emission \citep{smith02,smith06}.  As noted in the Introduction, \citet{currie02} referred to this structure as the ``Ghost shell''.  
This structure does not appear to extend spatially beyond the border of the SE lobe, and its velocity structure suggests that it is a thin bubble.  
It is therefore not coincident with the \mgii\, emission structures that are seen to extend more than 5{\arcsec} outside the polar lobes in images, and cover a wide range of radii (thus suggesting a range of expansion speeds in a Hubble-like flow).  
Nevertheless, both structures are outside the SE polar lobe and inside the radius of the ragged [N~{\sc ii}]-emitting shell seen as the S Ridge and S condensation, so they may be physically related.  
Based on its velocity structure and spatial location, the Ghost Shell appears to be limited to a smaller range of radii at the the outer boundary of the SE polar lobe, perhaps related to the dark edge of the SE lobe seen in \mgii.  
Long-slit UV spectra of this faint \mgii\, emission outside the polar lobes are needed to further investigate its velocity structure and possible kinematic relationship to the Ghost Shell.

\subsection{Equatorial Mg~II Emission}

Another dramatic feature seen in \mgii\, emission is the revelation of complete, possibly organized, jet-like features in the equatorial debris.  
The NN Jet has long been interpreted as a fast-moving column of material with an associated bow shock (e.g., \citealt{meaburn93}). 
Its core structure appears somewhat disembodied at its base in [N~{\sc ii}] F658N images, and even more ragged in F336W and other continuum-dominated images. 
However, the \mgii\, emission from the NN Jet core is brightest at the base and can be traced all the way back to the Homunculus equatorial region as a smooth and contiguous beam of UV light.

Likewise, the S Condensation on the opposite side of the Homunculus from the NN Jet has long been recognized as a shock interface between fast ejecta and circumstellar material \citep{dufour99}.  
While there were faint indications of connected material in the [N~{\sc ii}] and blue continuum images back toward the central star \citep{morse98}, the \mgii\, emission again shows a bright, contiguous column of material connecting the outer portions of the S Condensation all the way back to the Homunculus equatorial skirt.  
Moreover, there is another well-defined thin column of light just south of the S Condensation column, rotated by $\sim$15$^{\circ}$.   
These two radial beams of UV light seem to outline a dense clump of shadowing dust in the equatorial skirt, located 4{\arcsec} west and 2{\arcsec} south of the central star.

Thus the \mgii\, data show emphatically that the NN Jet and S Condensation are the same phenomenon, characterized observationally by a contiguous beam of resonantly scattered \mgii\, emission connecting to fast-moving, shocked condensations in the Outer Ejecta.  
These appear to be very special directions in the equatorial plane.   

Do these radial UV beams appear because that is where dense material lies in the outer parts of an equatorial skirt, or do we see them because they are akin to searchlights, extending from a hole at smaller radii where UV light escapes?  
Their smooth and perfectly radial structure seems to imply the latter, as with the streaked \mgii\, emission over the polar lobe regions outside the Homunculus.   
On the other hand, there are indeed dense, shocked gas structures at the terminus of each of the beams of UV light, and the NN Jet even includes a well-defined bow-shock structure that surrounds the end of that UV beam.  
The clumpy, dense [N~{\sc ii}]-emitting structure in the same directions indicate that there is indeed a dense outflow of fast-moving material here.  
Taken together, the fast, shocked [N~{\sc ii}]-emitting knots in the S Condensation and NN jet, and the \mgii\, beams, may suggest that dense knots of fast-moving gas blasted through the equatorial torus  in these directions, leaving a tunnel in their wake through which radial beams of near-UV light can now escape.  The escaping beams of UV light require that whatever may have punched these holes in the equatorial torus, the prominent holes in these special directions have persisted since the Great Eruption.

In fact, the dense equatorial waist of the Homunculus is seen at long mid-IR to submm wavelengths to be a disrupted and clumpy toroidal structure \citep{chesneau05,smith02c,smith05,smith18c}, making it likely that UV starlight is blocked in most azimuthal directions, but may escape between clumps in certain preferred directions.  
The current surviving companion to $\eta$ Car is on a very eccentric orbit with a known orientation in 3-D space \citep{madura12}.  
Interestingly, it has been noted that these two directions of the S Condensation and NN Jet, respectively, correspond to directions where this companion star might have plunged into and out of a disk around the erupting star \citep{smith18b,smith18c}, if such a disk were responsible for pinching the waist of the Homunculus during the Great Eruption. 

We note that there is also faint, diffuse \mgii\, emission located outside the Homunculus at locations that seem to coincide with latitudes within very roughly $\pm$45$^{\circ}$ above and below the equatorial plane, corresponding to mid-latitudes of the Homunculus.  As with the \mgii\, in the polar directions, this diffuse emission is bounded at its inner edge by the Homunculus and bounded on its outer edge by the bright [N~{\sc ii}] shell.   It is, however, much fainter than the emission near the polar axes.

\subsection{Mg~II emission, shocked [N~II] emission, and X-rays}

Figure~\ref{fig:xray} shows the same 3-color {\it HST} image from Figure~\ref{fig:color}, but superposed upon the extended X-ray emission from $\eta$~Car detected by the {\it Chandra X-ray Observatory}, with the X-ray emission shown in purple.  Although this image includes 0.5-7 keV flux,  most of the extended X-ray emission is concentrated in the softer X-ray band at 0.5-1 keV \citep{seward01,corcoran04,hamaguchi14}.

The main shell of extended X-ray emission marks a shock front between the fastest N-rich ejecta and an ancient pre-euption wind.  N abundances are higher inside X-ray shell and lower outside it \citep{sm04}.  Moreover, the fastest Doppler shifts in the Outer Ejecta are seen from diffuse N-rich filaments inside the soft X-ray shell that can only be detected in spectra \citep{smith08}, whereas the dense knots and clumps seen in F658N images with {\it HST} are slower \citep{mehner16,weis01,weis99} and arise from eruptions 300 or 600 years before the Great Eruption \citep{kiminki16}.  This suggests that the main soft X-ray shell is truly a blast wave from the Great Eruption, making the shocked shells surrounding $\eta$~Car somewhat analogous to a low-energy SNR.  The kinematics and geometry of the Outer Ejecta are complicated, originating from a series of previous eruptions.  \citet{kiminki16} and \citet{mehner16} have recently discussed the structure and kinematics of the Outer Ejecta in detail.

Since the main X-ray shell seen at relatively large radii is now crashing into ejecta from previous eruptions, one can assume that it has largely cleared away the older material that may have been located closer to the star in the past.  
The vast majority of ejecta inside the X-ray shell was ejected more recently, either during the main event of the 19th century Great Eruption \citep{smith17,morse01}, in the decades leading up to that eruption \citep{kiminki16}, or minor eruptions afterward \citep{smith04a,dorland04}.  
As noted earlier, the \mgii-emitting structures discovered here are sandwiched in between the Homunculus (with a well-determined ejection date of 1847) and the bright [N~{\sc ii}]-emitting Outer Ejecta in the S Ridge and similar features, which have dates of origin from proper motions that are either associated with the Great Eruption or within decades before the 1840s \citep{kiminki16}.  
Moreover, there is a clear deficit of X-ray emission at the locations of the brightest \mgii,\ emission itself, indicating that there is not a strong gradient in the velocity or characteristic age of material at this position (i.e., the \mgii\, emission arises from gas in free expansion, and does not arise at a shock interface where fast young material collides with slower older material).  
We therefore conclude that the \mgii-emitting structures were most-likely ejected by the star either in the main event of the Great Eruption, or perhaps in the decades leading up to it.  
If so, its expansion speed away from the star should be similar to or somewhat faster than the expansion of the polar lobes of the Homunculus \citep{smith06}, moving at 650 km~s$^{-1}$.  
This would suggest that the \mgii\, structures are moving at roughly 700$-$1000 km s$^{-1}$ based on their location.  
Proper motions from a second epoch of F280N imaging and/or Doppler velocities from \mgii\, $\lambda$2800 spectra are needed to more tightly constrain the speed and time of origin of the \mgii, features.  
The kinematics of this material could be of great interest if \mgii\, structures probe the polar mass-loss of $\eta$ Car in the decades leading up to the Great Eruption.  
This may offer an important diagnostic of the star system's growing instability, especially in the context of a possible merger event \citep{smith18b,pzvdh16}.

\subsection{Another connection to SLSNe}

The violent, impulsive mass ejection exemplified by $\eta$~Carinae in its 19th century eruption has been compared to the extreme pre-SN mass-loss of some SNe that show signatures of very strong shock interaction with circumstellar material, including SNe of Types IIn and Ibn (see \citealt{smith14}).  
Most notable among these are the cases of several super-luminous SNe of Type IIn, such as the well studied cases of SN~2006gy and SN~2005tf \citep{smith07,smith+08}.  
The physical parameters of their impulsive pre-SN mass loss events include massive shells ($\sim$20 $M_{\odot}$) ejected at speeds of hundreds of km~s$^{-1}$, for which the only known precedent in nearby stellar populations is the Great Eruption of $\eta$~Car.

In a recent study of the Type Ic super-luminous supernova iPTF16eh, \citet{lunnan18} discovered an unusual \mgii\, resonance light echo associated with the event.  
From the evolution of this \mgii\, echo, those authors inferred the existence of a massive, detached (0.1 pc) circumstellar shell ejected in an impulsive mass-loss event decades before the SN.  
Interestingly, they drew some comparisons to the nebula around $\eta$~Carinae, although there are differences as well (for instance, those authors infer a spherical geometry for the shell, unlike the bipolar ejecta around $\eta$~Car).  
Spectra of iPTF16eh indicated that the SN photosphere itself was hydrogen poor \citep{lunnan18}, but the H abundance in the detached shell is not yet known because so far it is only seen in resonance scattering of \mgii; its composition might be constrained by forthcoming observations obtained if/when the fast SN ejecta overtake and shock the detached shell.   
Although we would not necessarily expect the \mgii-emitting structures around $\eta$~Car to maintain the same level of ionization if $\eta$~Car were to explode as a super-luminous SN, we nevertheless found it noteworthy that both $\eta$~Car and an SLSN have dense shells seen in the unique tracer of near-UV \mgii\, resonant scattering.

\subsection{Mass of unshocked, neutral gas from Mg~II emission}

Since the structures seen here for the first time in resonant scattering of \mgii~$\lambda$2800 are not seen in any other previous imaging of $\eta$~Car, and are presumably dominated by neutral atomic gas, we cannot estimate their mass via normal diagnostics like thermal-IR dust emission or molecular emission, nor via recombination or collisionally excited emission from ionized plasma. 
We can, however, make a crude guess at the gas mass based on the \mgii\, emission itself and the assumption of resonant scattering.  
Future spectroscopy of the \mgii\, doublet can constrain the line scattering optical depth as well as the true expansion velocity of the gas. 
But for now we can make the rough assumption that for resonant scattering to be efficient, the optical depth of the brighter line at 2796 \AA \, is probably close to unity, i.e., $\tau_{2796} \approx 1$.  Following \citet{martin13}, who made a similar assumption in an analysis of the \mgii\, scattering halos of galaxies, the Sobolev optical depth can be written as

\begin{equation}
\tau_{2976} = 4.6 \times 10^{-7} {\rm cm}^{3} {\rm s}^{-1} \, \, n_{Mg^+} \big{(} \frac{dv}{dr} \big{)}^{-1}
\end{equation}

\noindent where $dv/dr$ is the velocity gradient at the location where the line scattering optical depth reaches unity, and which we assume is constant in this case with a Hubble-like flow.  
Adopting representative values of $v$=700 km s$^{-1}$ for the expansion speed and $r$=22,000 AU for the outer radius of the polar lobes of the Homunculus \citep{smith06}, and setting $\tau_{2796}$=1, we have

\begin{equation}
n_{Mg^+} \simeq 5 \times 10^{-4} {\rm cm}^{-3} \, \big{(}\frac{v}{700 km s^{-1}}\big{)} \big{(}\frac{r}{22,000 AU} \big{)}^{-1}
\end{equation}

\noindent for the number density of \mgii\, scattering atoms, which may be a conservative lower limit if there is additional mass hidden in very optically thick clumps. To convert this to a total H gas density, we need to correct for the abundance of Mg, $n_{Mg}/n_H$, the ionization fraction of Mg, $\chi = n_{Mg^+}/n_{Mg}$, and the fraction of the total Mg atoms that are left in the gas phase and not depleted onto dust grains, $f_g$ = $n_{Mg}$(gas) / $n_{Mg}$(total), i.e.,

\begin{equation}
n_H = n_{Mg^+} \Big{(} \frac{n_{Mg}}{n_H} \, \chi \, f_g  \Big{)}^{-1}.
\end{equation}

\noindent To be conservative, we adopt $\chi$=1, although some fraction of the Mg may be Mg$^0$ or Mg$^{++}$.  
We assume a solar abundance of Mg, roughly $n_{Mg}/n_H$ = 4$\times$10$^{-5}$.  
Unfortunately, for the goal of using Mg emission to constrain the mass, the large condensation temperature of Mg around 1300 K means that a large fraction of Mg atoms are typically depleted onto dust grains (leaving a small value of $f_g$ for Mg atoms in the gas phase), requiring a large and uncertain correction factor.  
From studies of the warm and cool atomic gas in the Galactic ISM, \citet{ss96} estimate values for Mg in the gas phase of roughly 3\% to 13\%.  Adopting the larger gas fraction for a conservative estimate, we therefore find an approximate density of $n_H \approx 100$ cm$^{-3}$.  
This may be an underestimate if there are clumps with high optical depth, or if more significant Mg depletion onto grains leaves a lower atomic gas fraction of Mg.

To estimate the total mass of the \mgii\, resonance scattering shell around $\eta$ Car, we have $M = m_H n_H V$, where $V = (4/3) \pi (R_{out}^3 - R_{in}^3)$ is the volume of a hollow spherical shell.  Taking $R_{in}$=22,000 AU as the inner radius (matching the outer extent of the Homunculus lobes at the pole; \citealt{smith06}) and $R_{out}$=30,000 AU as an approximate outer radius, the derived H density would imply a total mass of roughly 4$\times$10$^{31}$ g or about 0.02~$M_{\odot}$.  
Expanding with a typical speed of around 800 km s$^{-1}$, this mass has a total kinetic energy of around 1.3$\times$10$^{47}$ erg.  
We caution that these are very rough, order of magnitude estimates, although we also noted reasons why they may be conservative underestimates if wrong.  
These values are very small compared to the total mass and kinetic energy budget of the Great Eruption of around 15 $M_{\odot}$ and 10$^{50}$ erg \citep{smith03a,smith06,smith08}.  
They are comparable, however, to values estimated for the mass loss of the 1890 Lesser Eruption, which was roughly 0.1 $M_{\odot}$ and 10$^{46.9}$ erg \citep{smith05,bish03}.

Putting these mass and kinetic energy estimates in context requires more information about the origin of the gas in the \mgii\, scattering halo.  
Proper motions and radial-velocity structure in long-slit spectra are needed for this.  Residing outside the Homunculus polar lobes, it is interesting to speculate that this material might provide some clues about the mass-loss and instability of the star in the decades leading up to the peak of the Great Eruption.  
Spectroscopy of light echoes \citep{rest12,prieto14,smith18a,smith18b} shows relatively low outflow speeds of 150-200 km s$^{-1}$ in the early phases of the eruption.  
However, these echoes view $\eta$ Car from a vantage point near the equatorial plane.  If there was a dense and slow equatorial outflow in early phases, as expected in the scenario of a binary merger \citep{smith18b}, this slow and dense outflow would likely dominate spectra, but does not preclude a somewhat less dense and faster wind in the polar direction.  A polar wind might be expected if the pre-eruption star was rapidly rotating \citep{owocki96,og97,owocki98,do02}.
The \mgii-emitting gas reported here may therefore provide important and unique clues about $\eta$ Car's early mass loss and instability.  
Alternatively, the \mgii-emitting gas may have some other origin, such as material that leaked through the clumpy polar lobes due to hydrodynamic instabilities.



\section{SUMMARY}

We present the first {\it HST}/WFC3 UVIS and IR images of $\eta$~Carinae.  The most significant new result is that a deep image in the F280N filter, never used before to image $\eta$~Carinae, shows extended nebular structures outside the bipolar Homunculus nebula that have not been visible in any previous continuum or emission-line images.  The F280N filter includes \mgii\, $\lambda\lambda$2796,2803 emission that traces low-ionization atomic gas (mostly neutral H, low ionization metals).  It arises from a transition zone that appeared as a void in previous images, coming from an apparent cavity outside the molecular Homunculus polar lobes, but inside the shock-ionized Outer Ejecta.  We note a very striking radially streaked morphology to the \mgii\, emission, and we suspect that these arise from radial beams of UV light escaping the clumpy Homunculus.  In some cases, dark spaces between these radial streaks connect back to dense dust clumps in the polar lobes or equatorial skirt, as if they are long shadows projecting out into the regions outside the Homunculus.

While neutral atomic gas is hard to detect because it is invisible in most tracers, it may show up well in resonance scattering lines near a strong light source.  This fact, plus the observed \mgii\,  morphology (including pronounced radial streaks) and an anticorrelation with emission from shocked clumps, leads us to propose that resonance scattering is the dominant emission mechanism giving rise to the F280N emission outside the Homunculus that we detect.  We derive a mass of around 0.02 $M_{\odot}$ or more, and an outflow kinetic energy around 10$^{47}$ erg for the gas traced by \mgii.  This is small compared to the total mass and energy budget of the Great Eruption, although this estimate required assumptions that were chosen to be on the conservative side.  Thus, the quoted mass and energy might be underestimates, perhaps by as much as an order of magnitude.  Nevertheless, the \mgii\, emission may be tracing a component of the outflow that provides important clues to the early phases of the Great Eruption before the main Homunculus polar lobes were ejected, thus providing information about the building instability of the star.  These mass and energy estimates, and constraints on the date of origin for the material, can be improved with future {\it HST} spectroscopy and proper motions of the \mgii\, emission.

To our knowledge, no other LBVs have been observed with deep F280N imaging, so we do not know if the \mgii\, nebula around $\eta$ Carinae is unique.   We suspect, though, that resonance scattering of \mgii\, might be a good tracer of neutral or low-ionization gas in a variety of stellar outflows where warm gas is bathed in near-UV light, but where the central star might not be hot enough to fully ionize the outflow or not cool enough (or the outflow is not dense enough) to allow all the CSM to condense into dust and molecules.    One example already mentioned is the case of a \mgii\, resonance scattering nebula around a distant super-luminous SN \citep{lunnan18}.  Another example is the outflows from protostellar objects, where a shock-ionized component of the outflow is seen in visible-wavelength emission lines as Herbig-Haro (HH) jets, and where molecular outflows are seen at IR and radio wavelengths.  Comparing the two, molecular envelopes generally show a substantial excess of momentum as compared to the ionized gas in the collimated jets or bow shocks \citep[e.g.][]{dg16}, suggesting a low ionization fraction in the jet outflow and a larger reservoir of atomic gas.  Indeed, comparing H$\alpha$ to tracers of low-ionization gas reveals that the ionized gas emitting H$\alpha$ can be only a small portion of the jet outflow \citep{reiter15,reiter17}.  Perhaps \mgii\, resonance scattering would be a useful way to trace this low ionization gas in cases that do not suffer too much foreground extinction.

\section*{Acknowledgements}

\scriptsize 


We acknowledge helpful comments from the referee, which helped to clarify
several aspects of the observations and analysis presented in this paper.
We thank STScI science and technical staff members Peter McCullough and Amber Armstrong
for special assistance in planning the DD time observations and 
Joe DePasquale for producing the color 
{\it HST} image in Figure~\ref{fig:color} and the similar image
superposed on the {\it Chandra} X-ray image in Figure~\ref{fig:xray}.  
We thank both Kenji Hamaguchi and Mike Corcoran for discussions over many 
years concerning the X-ray emission from $\eta$ Car's outer shell. 
JM appreciates helpful discussions with John Raymond and Pat Hartigan.

Based on observations made with the NASA/ESA Hubble Space Telescope,
obtained at the Space Telescope Science Institute, operated by the
Association of Universities for Research in Astronomy, Inc., under
NASA contract NAS 5-26555.  Support was provided by NASA through
grants GO-15596, GO-15289, GO-14768, and AR-14586 from the Space
Telescope Science Institute, which is operated by the Association of
Universities for Research in Astronomy, Inc., under NASA contract NAS
5-26555. NS's research on Eta Carinae also received support from NSF
grants AST-1312221 and AST-1515559.


\begin{thebibliography}{}

\bibitem[Aghakhanloo et al.(2017)]{mojgan17} Aghakhanloo M, Murphy J,
  Smith N, Hlozek R. 2017, MNRAS, 472, 591

\bibitem[Blair et al.(2000)]{blair00} Blair WP, Morse JA, Raymond JC, 
Kirshner RP, Hughes JP, Dopita MA, Sutherland RS, Long KS, Winkler PF. 
2000, ApJ, 537, 667

\bibitem[Chesneau et al.(2005)]{chesneau05} Chesneau O. et al.\ 2005,
  A\&A, 435, 104


\bibitem[Chevalier \& Kirshner(1978)]{chev78} Chevalier, R.A., Kirshner, R.P. 1978, ApJ, 219, 931

\bibitem[Chevalier \& Liang(1989)]{chev89} Chevalier, R.A., Liang, E.P. 1989, ApJ, 344, 332

\bibitem[Corcoran et al.(2004)]{corcoran04} Corcoran, M. F., Hamaguchi, K., Gull, T., et al. 2004, ApJ, 613, 381

\bibitem[Currie et al.(1996)]{currie96} Currie, D. G., et al. 1996,
  AJ, 112, 1115

\bibitem[Currie et al.(2002)]{currie02} Currie, D. G., et al. 2002 - ghost shell

\bibitem[Damineli(1996)]{damineli96} Damineli, A. 1996, ApJ, 460, L49



\bibitem[Davidson et al.(1982)]{davidson82} Davidson K., Walborn
  N. R., Gull T. R., 1982, ApJ, 254, L47



\bibitem[Davidson et al.(2001)]{davidson01} Davidson, K., Smith, N.,
  Gull, T. R., Ishibashi, K., \& Hillier, D. J. 2001, AJ, 121, 1569

\bibitem[Dionatos \& G\"udel(2016)]{dg16} Dionatos O., G\"udel M., 2016, A\&A, 597, A64

\bibitem[Dorland et al.(2004)]{dorland04} Dorland, B. N., Currie,
  D. G., \& Hajian, A. R. 2004, AJ, 127, 1052

\bibitem[Dufour et al.(1999)]{dufour99} Dufour RJ, Glover TW, Hester
  JJ, Currie DG, van Orsow D, Walter DK. 1999, in ASP
  Conf.\ Ser.\ 179, Eta Carinae At The Millennium, ed. JA Morse et
  al. (San Francisco: ASP), 134
  
\bibitem[Dwarkadas \& Owocki(2002)]{do02} Dwarkadas VV, Owocki SP. 2002, ApJ, 581, 1337
  


\bibitem[Gaviola(1950)]{gaviola50} Gaviola, E. 1950, ApJ, 111, 408







\bibitem[Gull et al.(2005)]{gull05} Gull TR, Viera G, Bruhweiler F,
  Nielsen KE, Verner E, Danks A.  2005, ApJ, 620, 442

\bibitem[Gull et al.(2006)]{gull06} Gull TR, Kober GV, Nielsen
  KE. 2006, ApJS, 163, 173

\bibitem[Hamaguchi et al.(2014)]{hamaguchi14} Hamaguchi K., Corcoran M. F., 
Russell C. M. P. et al., 2014, ApJ, 784, 125

\bibitem[Hartigan et al.(1987)]{hartigan87} Hartigan, P., Raymond, J., 
  Hartmann, L. 1987, ApJ, 316, 323

\bibitem[Hartigan et al.(2004)]{hartigan04} Hartigan, P., Raymond, J., 
  Pierson, R. 2004, ApJ, 614, L69

\bibitem[Herschel(1847)]{herschel1847} Herschel J.F.W.\ 1847, Results
  of Astronomical Observations Made during the Years 1834, 5, 6, 7, 8
  at the Cape of Good Hope (London: Smith, Elder \& Co.)

\bibitem[Hester et al.(1991)]{hester91} Hester JJ, Light RM, Westphal
  JA, Currie G, Groth EJ, Holtzmann JA, Lauer TR, O'Neil EJ.\ 1991,
  AJ, 102, 654

\bibitem[Hillier \& Allen(1992)]{ha92} Hillier, D. J., \& Allen,
  D. A. 1992, A\&A, 262, 153


\bibitem[Hillier et al.(2006)]{hillier06} Hillier, D. J., Gull, T.,
  Nielsen, K., et al. 2006, ApJ, 642, 1098 


\bibitem[Ishibashi et al.(2003)]{bish03} Ishibashi K. et al., 2003, AJ, 125, 3222



\bibitem[Kiminki et al.(2016)]{kiminki16} Kiminki MM, Reiter M, Smith
  N. 2016, MNRAS, 463, 845



\bibitem[Lunnan et al.(2018)]{lunnan18} Lunnan R, Fransson C,
  Vreeswijk PM, et al. 2018, Nature Astronomy, 2, 887

\bibitem[Madura et al.(2012)]{madura12} Madura T.I., et al. 2012,
  MNRAS, 420, 2064

\bibitem[Martin et al.(2013)]{martin13} Martin CL, Shapley AE, Coil AL, 
Kornei KA, Murray N, Pancoast A. 2013, ApJ, 770, 41

\bibitem[Meaburn et al.(1993)]{meaburn93} Meaburn, J., Gehring, G.,
Walsh, J.R., Palmer, J.W., Lopez, J.A., Bryce, M., Raga, A.C. 1993,
A\&A, 276, L21

\bibitem[Mehner et al.(2016)]{mehner16} Mehner A, et al.\ 2016, A\&A,
  595, A120  


\bibitem[Morse et al.(1998)]{morse98} Morse, J. A., Davidson, K.,
  Bally, J., Ebbets, D., Balick, B., Frank, A. 1998, AJ, 116, 2443

\bibitem[Morse et al.(2001)]{morse01} Morse J. A., Kellogg J. R.,
  Bally J., Davidson K., Balick B., Ebbets D., 2001, ApJ, 548, L207

\bibitem[Nielsen et al.(2005)]{nielsen05} Nielsen K. E., Gull T. R.,
  Viera Kober G., 2005, ApJS, 157, 138

\bibitem[Owocki \& Gayley(1997)]{og97} Owocki SP, Gayley KG. 1997, in
  Luminous Blue Variables: Massive Stars in Transition, ASP Conf.\
  Ser. 120, ed. A.\ Nota \& H.\ Lamers (San Francisco: ASP), 121

\bibitem[Owocki et al.(1996)]{owocki96} Owocki SP, Cranmer SR, Gayley KG. 1996, ApJ, 472, L115

\bibitem[Owocki et al.(1998)]{owocki98} Owocki SP, Gayley KG, Cranmer SR. 1998, in ASP Conf. Ser.
131, Boulder Munich II: Properties of Hot Luminous Stars, ed. I. D.
Howarth (San Francisco: ASP), 237


\bibitem[Owocki et al.(2004)]{owocki04} Owocki S. P., Gayley K. G., Shaviv N. J., 2004, ApJ, 616, 525




\bibitem[Portegies Zwart \& van den Heuvel(2016)]{pzvdh16} Portegies
  Zwart S. F., van den Heuvel E. P. J., 2016, MNRAS, 456, 3401

\bibitem[Prieto et al.(2014)]{prieto14} Prieto JL, et al.\ 2014, ApJ,
  787, 8



\bibitem[Reiter et al.(2015)]{reiter15} Reiter M, Smith N, Kiminki MM, Bally J. 2015, MNRAS, 450, 564

\bibitem[Reiter et al.(2017)]{reiter17} Reiter M, Kiminki MM, Smith N, Bally J. 2017, MNRAS, 470, 4671

\bibitem[Rest et al.(2012)]{rest12} Rest A, et al.\ 2012, Nature, 482,
  375






\bibitem[Savage \& Semback(1996)]{ss96} Savage B, Semback K. 1996, ARA\&A, 34, 279

\bibitem[Seward et al.(2001)]{seward01} Seward, F. D., Butt, Y. M.,
  Karovska, M., Prestwich, A., Schlegel, E. M., Corcoran, M. F. 2001,
  ApJ, 553, 832

\bibitem[Shaviv(2000)]{shaviv00} Shaviv NJ. 2000, ApJ Letters, 532,
  L137



\bibitem[Smith(2002a)]{smith02} Smith N.\ 2002a, MNRAS, 337, 1252
  
\bibitem[Smith(2002b)]{smith02b} Smith N.\ 2002b, MNRAS, 336, L223

\bibitem[Smith(2004)]{smith04} Smith N.\ 2004, MNRAS, 351, L15

\bibitem[Smith(2005)]{smith05} Smith N.\ 2005, MNRAS, 357, 1330

\bibitem[Smith(2006)]{smith06} Smith N.\ 2006, MNRAS, 644, 1151 

\bibitem[Smith(2008)]{smith08} Smith N.\ 2008, Nature, 455, 201



\bibitem[Smith(2013)]{smith13} Smith N.\ 2013, MNRAS, 429, 2366

\bibitem[Smith(2014)]{smith14} Smith N.\ 2014, ARAA, 52, 487

\bibitem[Smith(2017)]{smith17} Smith N.\ 2017, MNRAS, 471, 4465 


\bibitem[Smith \& Davidson(2001)]{sd01} Smith N., Davidson K., 2001, ApJ, 551, L101


\bibitem[Smith \& Ferland(2007)]{sf07} Smith N., Ferland G. J., 2007, ApJ, 655, 911

\bibitem[Smith \& Frew(2011)]{sf11} Smith N, Frew D.\ 2011, MNRAS, 415, 2009

\bibitem[Smith \& Hartigan(2006)]{sh06} Smith N, Hartigan P. 2006, ApJ, 638, 1045
  

\bibitem[Smith \& Morse(2004)]{sm04} Smith N, Morse JA. 2004, ApJ,
  605, 854

\bibitem[Smith \& Owocki(2006)]{so06} Smith N, Owocki SP, 2006,
  ApJ, 645, L45


\bibitem[Smith \& Tombleson(2015)]{st15} Smith N, Tombleson R. 2015,
  MNRAS, 447, 602
 
\bibitem[Smith et al.(2000)]{smith00} Smith, N., Morse, J. A.,
  Davidson, K., Humphreys, R. M. 2000, AJ, 120, 920 L145

\bibitem[Smith et al.(2002)]{smith02c} Smith N., Gehrz R. D., Hinz
  P. M., Hoffmann W. F., Mamajek E. E., Meyer M. R., Hora J. L.  2002,
  ApJ, 567, L77

\bibitem[Smith et al.(2003a)]{smith03a} Smith N, Davidson K, Gull TR,
  Ishibashi K, Hillier DJ.  2003a, ApJ, 586, 432

\bibitem[Smith et al.(2003b)]{smith03b} Smith N., Gehrz R. D., Hinz
  P. M., Hoffmann W. F., Hora J. L., Mamajek E. E., Meyer M. R., 2003b,
  AJ, 125, 1458

\bibitem[Smith et al.(2004a)]{smith04a} Smith N., Morse, J.A., Gull
  T.R., et al., 2004a, ApJ, 605, 405

\bibitem[Smith et al.(2004b)]{smith04b} Smith N., Morse J. A., Collins
  N. R., Gull T. R., 2004b, ApJ, 610, L105

\bibitem[Smith et al.(2007)]{smith07} Smith N., Li W., Foley RJ, et al. 2007, ApJ, 666, 1116

\bibitem[Smith et al.(2008)]{smith+08} Smith N., Chornock R., Li W., et al. 2007, ApJ, 686, 467

\bibitem[Smith et al.(2011)]{smith+11} Smith N., et al.\ 2011, MNRAS,
  415, 773 



\bibitem[Smith et al.(2018a)]{smith18a} Smith N, Rest A, Andrews JE,
  Matheson T, Bianco FB, Prieto JL, James DJ, Smith RC, Strampelli GM,
  Zenteno A. 2018a, MNRAS, 480, 1457

\bibitem[Smith et al.(2018b)]{smith18b} Smith N, Andrews JE, Rest A,
  Bianco FB, Prieto JL, Matheson T, James DJ, Smith RC, Strampelli GM,
  Zenteno A. 2018b, MNRAS, 480, 1466
  
\bibitem[Smith et al.(2018c)]{smith18c} Smith N., Ginsburg A, Bally
  J. 2018, MNRAS, 474, 4988

\bibitem[Steffen et al.(2014)]{steffen14} Steffen W, Teodoro M, Madura
  TI, et al.\ 2014, MNRAS, 442, 3316



\bibitem[Sutherland \& Dopita(1995)]{sd95} Sutherland RS, Dopita MA. 1995, ApJ, 439, 381

\bibitem[Teodoro et al.(2008)]{teodoro08} Teodoro M., Damineli A.,
  Sharp R. G., Groh J. H., Barbosa C. L., 2008, MNRAS, 387, 564








\bibitem[Van Dyk \& Matheson(2012)]{vdm12} Van Dyk SD, Matheson
  T. 2012, in Eta Carinae and the Supernova Impostors,
  ed. R.M. Humphreys and K. Davidson
  (Springer) 


\bibitem[van Marle et al.(2008)]{vanmarle08} van Marle AJ, Owocki SP,
  Shaviv NJ.\ 2008, MNRAS, 389, 1353

\bibitem[van Marle et al.(2009)]{vanmarle09} van Marle AJ, Owocki SP,
  Shaviv NJ.\ 2009, MNRAS, 394, 595

\bibitem[Verner et al.(2005)]{verner05} Verner E., Bruhweiler F.,
  Nielsen K. E., Gull T. R., Viera kober G., Corcoran M. F., 2005,
  ApJ, 629, 1034

\bibitem[Walborn(1976)]{walborn76} Walborn N. R., 1976, ApJ, 204, L17


\bibitem[Walborn et al.(1978)]{walborn78} Walborn, N. R., Blanco, B. M., 
Thackerey, A. D. 1978, ApJ, 219, 498


\bibitem[Weigelt \& Kraus(2012)]{wk12} Weigelt G, Kraus S. 2012, ASSL, 384, 129

\bibitem[Weis(2012)]{weis12} Weis K., 2012, in Davidson K., Humphreys
  R. M., eds, Astrophysics and Space Science Library,
  Vol. 384. Astrophysics and Space Science Library.  p. 171

\bibitem[Weis et al.(1999)]{weis99} Weis K., Duschl W. J., Chu
  Y.H. 1999, A\&A, 349, 467

\bibitem[Weis et al.(2001)]{weis01} Weis K., Duschl W. J., Bomans
  D. J., 2001, A\&A, 367, 566

\bibitem[Woosley(2017)]{woosley17} Woosley SE. 2017, ApJ, 836, 244

\end{thebibliography}

\end{document}